\documentclass[aps, prd, twocolumn, superscriptaddress, nofootinbib]{revtex4}
\usepackage{graphicx}% Include figure files
\usepackage{dcolumn}% Align table columns on decimal point
\usepackage{amssymb}% outlined letters
\usepackage{amsmath,amssymb,amsfonts}

\begin{document}

%Macros
\newcommand{\Eq}[1]{\mbox{Eq. (\ref{eqn:#1})}}
\newcommand{\Fig}[1]{\mbox{Fig. \ref{fig:#1}}}
\newcommand{\Sec}[1]{\mbox{Sec. \ref{sec:#1}}}

\newcommand{\PHI}{\phi}
\newcommand{\PhiN}{\Phi^{\mathrm{N}}}
\newcommand{\vect}[1]{\mathbf{#1}}
\newcommand{\Del}{\nabla}
\newcommand{\unit}[1]{\;\mathrm{#1}}
\newcommand{\x}{\vect{x}}
\newcommand{\y}{\vect{y}}
\newcommand{\p}{\vect{p}}
\newcommand{\ScS}{\scriptstyle}
\newcommand{\ScScS}{\scriptscriptstyle}
\newcommand{\xplus}[1]{\vect{x}\!\ScScS{+}\!\ScS\vect{#1}}
\newcommand{\xminus}[1]{\vect{x}\!\ScScS{-}\!\ScS\vect{#1}}
\newcommand{\diff}{\mathrm{d}}

\newcommand{\be}{\begin{equation}}
\newcommand{\ee}{\end{equation}}
\newcommand{\bea}{\begin{eqnarray}}
\newcommand{\eea}{\end{eqnarray}}
\newcommand{\vu}{{\mathbf u}}
\newcommand{\ve}{{\mathbf e}}

%=====================================================================
%=====================================================================
%=====================================================================

%\title{Quantization of fluctuations with de Sitter momentum space: the two-point function and beyond}
\title{Quantization of fluctuations in DSR: the two-point function and beyond}

\newcommand{\addressImperial}{Theoretical Physics, Blackett Laboratory, Imperial College, London, SW7 2BZ, United Kingdom}
\newcommand{\addressRoma}{Dipartimento di Fisica, Universit\`a ``La Sapienza''
and Sez. Roma1 INFN, P.le A. Moro 2, 00185 Roma, Italia}

\author{Giulia Gubitosi}
%\affiliation{\addressRoma}
\affiliation{\addressImperial}
\author{Michele Arzano}
\affiliation{\addressRoma}
\author{Jo\~{a}o Magueijo}
%\email{magueijo@ic.ac.uk}
\affiliation{\addressImperial}
%\affiliation{\addressRoma}

\date{\today}

\begin{abstract}
We show that the two-point function of a quantum field theory with de Sitter momentum space (herein called DSR) can be expressed as the product of a standard delta function and an energy-dependent factor. This is a highly non-trivial technical result in any theory without a preferred frame. Applied to models exhibiting running of the dimensionality of space, this result is essential in proving that vacuum fluctuations are {\it generally} scale-invariant at high energies whenever there is running to two dimensions. This is equally true  for theories with and without a preferred frame, with differences arising only as we consider higher order correlators. Specifically, the three-point function of DSR has a unique structure of ``open triangles'', as shown here. 
\end{abstract}


\maketitle

%=====================================================================
%=====================================================================
%=====================================================================

\section{Introduction}

Non-trivial properties of momentum space are a common feature emerging from several quantum gravity theories \cite{AmelinoCamelia:2008qg, AmelinoCamelia:2011pe, AmelinoCamelia:2011bm, Matschull:1997du, Freidel:2005me, Arzano:2014ppa, Gubitosi:2013rna}. An interesting general result \cite{vacfluctuations} is that whenever the dimensionality of energy-momentum space runs to two in the UV, then the power spectrum of (scalar and tensor) vacuum fluctuations is scale invariant. This is of great interest for cosmology, since it would provide an alternative mechanism to inflation for explaining the observed quasi-scale invariance of cosmological primordial fluctuations. The inclusion of gravity into this picture requires that it be conformally coupled to all matter in the UV, as shown explicitly for Horava-Lifschitz theories \cite{measure}, and speculated in the general case in~\cite{essay}. 
As explained in~\cite{vacfluctuations}  this  result  is valid for both Lorentz-violating theories and for relativistic theories which merely deform the Lorentz symmetries. 

The derivation in~\cite{vacfluctuations}  follows from the postulate that  {\it for the purpose of computing the two-point function}, quantization  can proceed from an undeformed commutator for creation and annihilation operators. This should  be set equal to a delta function adjusted to the deformed measure of integration in momentum space, but no further complexities are required. Such a postulate implicitly assumes that at the level of the two-point function the conservation rule of momenta can be expressed in the standard form, with a different energy-dependent coefficient multiplying the standard delta function. It is precisely this coefficient that renders the two-point function scale-invariant when momentum dimensionality runs to two in the UV. 

One might  fear that this quantization postulate surreptitiously introduces a preferred frame into theories purporting to be frame-independent. In such theories
frame invariance is achieved as a result of the concurring consistent deformation of a number of ingredients, namely the dispersion relation, the momentum-space integration measure and the composition rule of momenta in interactions. The last ingredient renders momentum space a non-abelian group manifold. This requires a deformation of the Fock space \cite{arzano-marciano, Arzano:2013sta}, with associated deformed commutators between creation and annihilation operators.  It would appear that our simplistic general quantization postulate contradicts the requirements imposed by frame independence.

In this paper we allay this concern, considering the concrete case of de Sitter momentum space coordinatized by the $\kappa$-Poincar\'e bicross-product basis \cite{Majid:1994cy}, herein labelled DSR (deformed special relativity). In Section~\ref{standard} we review the general quantization framework proposed in~\cite{vacfluctuations}. Then in Section~\ref{DSR} we derive the form of the general postulate as applied to DSR from the more rigorous quantization  of the theory. In Section~\ref{DSR1} we also perform this exercise in linearizing coordinates, helping the translation between the two frameworks. The complexities of DSR quantum field theory therefore do not affect the calculation of its two-point function, which falls within the general framework proposed in~\cite{vacfluctuations} for all theories. They do, however, come to the fore as we compute higher order correlators, as we show in Section~\ref{3point}.

\section{Standard scalar field quantisation with a deformed measure}\label{standard}

In this section we briefly review the standard quantisation of a scalar field under a deformed measure, in order to fix notation (see also~\cite{vacfluctuations}).
A scalar field can be expanded in positive and negative energy components in Fourier space according to:
\be
\phi(x)=\int d\mu(k) \left[\tilde\phi^{+}(k)e^{+}_{k}(x)+\tilde\phi^{-}(k)e^{-}_{k}(x)\right]\,,
\ee
where $e^{\pm}_{k}$ are positive (respectively, negative) energy plane waves, such that $e^{-}_{k}=(e^{+}_{k})^{\dagger}$. Here $d\mu(k)$ is the measure in four-dimensional momentum space.
Before quantising the field, one needs to go on-shell:
\be
\phi(x)=\int d\mu(k) \delta(\Omega)\theta(k_{0})  \left[\tilde\phi^{+}(k)e^{+}_{k}(x)+\tilde\phi^{-}(k)e^{-}_{k}(x)\right]\,,
\ee
where $\Omega$ is the mass-shell constraint. The above integral can be 	equivalently written as
\be
\phi(x)=\int d\bar\mu(\vec k)  \left[\tilde\phi^{+}(\omega_{k},\vec k)e^{+}_{k}(x)+\tilde\phi^{-}(\omega_{k},\vec k)e^{-}_{k}(x)\right]\,,
\ee
where $\omega_{k}$ is the positive solution to the mass-shell constraint and $d\bar\mu(\vec k)=\frac{d\mu(k)}{F(\omega_{k})}$ is the covariant on-shell measure on spatial momentum space (the function $F(\omega_{k})$ is the one resulting from the integration of the delta enforcing the mass-shell constraint). In the standard case, as is well known, we have:
 \be
d\bar\mu(\vec k)=\frac{d^{3}\vec k}{(2\pi)^{3}2\omega_{k}}.
\ee

Upon introducing creation and annihilation operators $a,a^{\dagger}$, such that $\left[a(\vec k) ,a^{\dagger}(\vec p) \right]=2\omega_{k}\delta^{(3)}(\vec k-\vec p)$, the field operator reads:
\be
\phi(x)=\int d\bar\mu(\vec k)  \left[a(\vec k) e^{+}_{k}(x)+a^{\dagger}(\vec k) e^{-}_{k}(x)\right]\,. \label{eq:fieldaexpansion}
\ee
 % 
% 
% 
% 
%The modes associated to on-shell plane wave are written as:
%\begin{equation}
%e_{\vec p}({\vec k})= 2 \omega ({\vec k}) \delta^{(3)}(\vec p- \vec k) 
%\end{equation}¥
%where $ \omega ({\vec k}) $ is the on-shell energy.
%Given these modes the inner product of two one-particle states is:
%\begin{equation}
%\left(e_{\vec k_{1}},e_{\vec k_{2}}\right)=\int \frac {d^{3} k}{2\omega(\vec k)} e_{\vec k_{1}}^{-}({\vec k})e_{\vec k_{2}}^{+}({\vec k})=2 \omega ({\vec k_{1}}) \delta^{(3)}(\vec k_{1}- \vec k_{2})
%\end{equation}
The one-particle state normalisation can be easily computed as:
\be
\langle k|k'\rangle  \equiv \langle 0 |   \left[a(\vec k),a^{\dagger}(\vec k')\right]|0\rangle  = 2 \omega_{k}\delta^{(3)}(\vec k-\vec k')\,.
\ee
We stress that  {\it this is precisely the normalization that will in general determine the power spectrum of quantum vacuum fluctuations}. 
Note that, using the notation in \cite{vacfluctuations}, we can equivalently write:
 \be
 \langle k|k'\rangle  \equiv \langle0 |   \left[a(\vec k),a^{\dagger}(\vec k')\right]|0\rangle  = \delta_{\bar\mu}(\vec k-\vec k')\,,
 \ee
where $\delta_{\bar\mu}(\vec p)$ is such that, for any function $f(\vec k)$:
\be
\int d\bar \mu(\vec p) \delta_{\bar\mu}(\vec p -\vec k) f(\vec p)=f(\vec k)\,.
\ee
The commutation rule of the creation and annihilation operators fixes the two-point correlation function, but 
this is only relevant at the level of the normalization of the one-particle state. Indeed the fluctuations power spectrum $P_{\phi}(\vec p)$ is given by:
 \be
 \langle 0|\phi(x)^{2}|0\rangle \equiv  \int \frac{d^{3}\vec p}{(2\pi)^3}P_{\phi}(\vec p)\,,
 \ee
 where the expansion (\ref{eq:fieldaexpansion}) has to be used in order to evaluate the l.h.s. Only
one particle states are involved in the calculation.

\section{Two-point function in relativistic de Sitter momentum space}\label{DSR}

The specific realisation of relativistic de Sitter momentum space that we are going to consider here is known as $\kappa$-Poincar\'e \cite{Majid:1994cy}. In terms of the so-called ``bicross-product" coordinates, the momentum space measure reads:
\be
d\mu(p)=e^{3p_{0}/\kappa}d^{3}\vec p\, dp_{0}.
\ee
As we mentioned in the introduction, relativistic compatibility requires that the on-shell relation and the composition rule of momenta are also consistently modified. In this case the on-shellness is given by:
\be
\mathcal C_{\kappa}=m^{2},
\ee
with 
\be
\mathcal C_{\kappa}\equiv 4 \kappa^{2}\sinh^{2}\left(p_{0}/2\kappa\right)+e^{p_{0}/\kappa}|\vec p|^{2}.
\ee
The composition rule for spatial momenta is:
\be
\vec p\oplus \vec q=\vec p+e^{-p_{0}/\kappa}\vec q\,, \label{eq:kappacomposition}
\ee
and this requires also a modified ``antipode", $\ominus p$, such that $p\oplus(\ominus p)=0$, and therefore
given by:
\be
 \ominus \vec p=-e^{p_{0}/\kappa}\vec p. \label{eq:kappaantipode}
\ee
The deformed on-shellness implies that the positive-energy (massless) solution is now:
\begin{equation}
\omega_{p}^{(\kappa)}=-\kappa \log\left(1-\frac{|\vec p|}{\kappa}\right) \,.\label{eq:omega}
\end{equation}
The deformed composition rule enters in the one-particle normalisation, as shown in \cite{MicheleQuantization, arzano-marciano}, according to expression:
\be
\langle k|k'\rangle   = 2 \omega_{k}^{(\kappa)}\delta^{(3)}(\vec k\oplus(\ominus \vec k'))\,, \label{eq:oneparticle}
\ee
where the $\delta^{(3)}(\cdot)$ is defined with respect to the deformed measure:
\be
\int d^{3}\vec p \,e^{3 p_{0}/\kappa}\;\delta^{(3)}(\vec p\oplus(\ominus \vec q)) f(\vec p)\equiv f(\vec q) \,.\label{eq:kappadelta}
\ee 
This normalisation is compatible with a deformed commutator for the creation and annihilation operators:
\be
\left[a(\vec k) ,a^{\dagger}(\vec p) \right]=2\omega_{k}\delta^{(3)}(\vec k\oplus(\ominus \vec p)), \label{eq:commutator}
\ee
which in turn  affects the two-point function, as shown in the previous section.

What is interesting, and is the main point of this paper, is that, as long as the above commutator is only evaluated on the vacuum, it reduces to the standard one, multiplied by a factor which is a function of energy.
The key point in order to show this is to observe that the equations (\ref{eq:oneparticle}) and (\ref{eq:commutator}) assume the on-shell condition.
In this case the deformed composition and antipode read:
\bea
\vec p\oplus \vec q&=&\vec p+e^{-\omega_{p}^{(\kappa)}/\kappa}\vec q \,,\\%= \vec p+\vec q\left(1-\frac{|\vec p|}{\kappa}\right), \\
  \ominus \vec p&=&-e^{\omega_{p}^{(\kappa)}/\kappa}\vec p\,,%=-\frac{\vec p}{1-\frac{|\vec p|}{\kappa}}          
\eea
so that the deformed delta function can be rewritten as:
\begin{eqnarray}
\delta^{(3)}(\vec k_{1}\oplus(\ominus \vec k_{2}))&=&\delta^{(3)}(\vec k_{1}- e^{-\omega_{k_{1}}^{(\kappa)}/\kappa}e^{\omega^{(\kappa)}_{k_{2}}/\kappa}\vec k_{2})\nonumber\\
%&=& \delta^{(3)}\left(e^{-\omega_{\kappa}(\vec k_{1})/\kappa} \left( e^{\omega_{\kappa}(\vec k_{1})/\kappa}\vec k_{1}- e^{\omega_{\kappa}(\vec k_{2})/\kappa}\vec k_{2}\right)\right)\\
&=&\delta^{(3)}(\vec k_{1}-\vec k_{2})e^{-\omega_{k_{1}}^{(\kappa)}/\kappa}\,, \label{eq:deltaredefinition}
\end{eqnarray}
where we used the fact that $\omega_{p}^{(\kappa)}$ is a function of $\vec p$ (see eq. (\ref{eq:omega})).
The one-particle normalisation is then:
\be
\langle k|k'\rangle   = 2 \omega_{k}^{(\kappa)}e^{-\omega_{k}^{(\kappa)}/\kappa}\delta^{(3)}(\vec k-\vec k')\,, 
\ee
and the commutator for the creation and annihilation operators:
\be
\left[a(\vec k) ,a^{\dagger}(\vec p) \right]=2\omega_{k}^{(\kappa)}e^{-\omega_{k}^{(\kappa)}/\kappa}\delta^{(3)}(\vec k-\vec k').
\ee
Recall that the delta function in this expression is not the standard three-dimensional delta function, since Eq. (\ref{eq:kappadelta}) must still be true. Indeed, when inserting the above result, Eq. (\ref{eq:deltaredefinition}), in (\ref{eq:kappadelta}), and remembering that everything is on-shell, we find:
\bea
&&\int d^{3}\vec p \,e^{3 \omega_{p}^{(\kappa)}/\kappa}\;\delta^{(3)}(\vec p\oplus(\ominus \vec q)) f(\vec p)\nonumber\\
&&=  \int d^{3}\vec p \,e^{2 \omega_{p}^{(\kappa)}/\kappa}\;\delta^{(3)}(\vec p-\vec q) f(\vec p)   \equiv f(\vec q).
\eea 
This gives  the relation between the above delta function and the standard three dimensional one, $\delta^{(3)}_{st}$:
\be
\delta_{st}^{(3)}(\vec p-\vec q)= e^{2 \omega_{p}^{(\kappa)}/\kappa} \;\delta^{(3)}(\vec p-\vec q).
\ee
In terms of the standard delta the one particle normalisation is:
\be
\langle k|k'\rangle   = 2 \omega_{k}^{(\kappa)}e^{-3\omega_{k}^{(\kappa)}/\kappa}\delta^{(3)}_{st}(\vec k-\vec k') \,,
\ee
and the commutator for the creation and annihilation operators:
\be
\left[a(\vec k) ,a^{\dagger}(\vec p) \right]=2\omega_{k}^{(\kappa)}e^{-3\omega_{k}^{(\kappa)}/\kappa}\delta^{(3)}_{st}(\vec k-\vec k').
\ee
We stress that this result is valid only when the commutator is evaluated on vacuum, and so affects only the two-point function. In fact it comes from the normalisation of the one particle state, which is only sensitive to this case. When considering multi-particle states, the resulting conservation rule of momenta will in general be modified. We will come back to this issue in section \ref{3point}.

\section{Two point function in linearising coordinates}\label{DSR1}
In \cite{vacfluctuations} we computed the two-point function and the power spectrum in terms of a different set of coordinates than the one used above. These were such that the on-shell relation takes standard form in the UV, so that all non-trivial features are encoded in the momentum space measure and the composition rule of momenta.
Such ``linearising'' coordinates are given by:
\be
\tilde p^{0}\equiv \kappa \, e^{p^{0}/2\kappa},\qquad \tilde p\equiv e^{p_{0}/2\kappa} p.
\ee
For consistency with the deformed summation rule (\ref{eq:kappacomposition}) and antipode, eq. (\ref{eq:kappaantipode}), the linearising variables satisfy:
\begin{eqnarray}
 \vec{ \tilde p}\oplus \vec{\tilde q}=\frac{\kappa^{2}}{(\tilde p^{0})^{2}}\left(\frac{(\tilde p^{0})^{2} }{\kappa^{2}} \vec{\tilde p}  +  \vec{\tilde q} \right)
\end{eqnarray}
and
\begin{eqnarray}
  \ominus\vec{ \tilde p} = - \left(\frac{\tilde p^{0}}{\kappa}\right)^{2}\vec{ \tilde p}~.
\end{eqnarray}
The positive-energy solution of the on-shellness is just the standard one:
\be
\omega_{\tilde k}=\tilde k^{0}\,.
\ee
Then the delta function appearing in the single-particle normalisation and in the creation and annihilation operators can be written as:
\begin{equation}
\delta^{(3)}(\vec{\tilde p}\oplus(\ominus \vec{\tilde k}))= \delta^{(3)}\left(\vec{\tilde p} - \frac{\omega_{\tilde k}^{2}}{\omega_{\tilde p}^{2}}\,\vec{ \tilde k}\right)=\frac{1}{3} \delta^{(3)}(\vec{\tilde p}- \vec{ \tilde k})\,.
\end{equation}
In analogy with what we have done in the bicross-product coordinates, we can write this in terms of a standard three-dimensional delta by observing that in this case
\be
\delta^{(3)}_{st}(\vec {\tilde p}- \vec {\tilde q}) =\frac{1}{3} \omega_{\tilde p}^{2}\,\delta^{(3)}(\vec {\tilde p}- \vec {\tilde q})\,.
\ee
The one-particle normalisation is therefore:
\be
\langle \tilde k|\tilde k'\rangle   = 2 \omega_{\tilde k}^{-1} \delta^{(3)}_{st}(\vec k-\vec k') 
\ee
and, similarly, the commutator for the creation and annihilation operators:
\be
\left[a(\vec{\tilde k}) ,a^{\dagger}(\vec{\tilde k}') \right]=2 \omega_{\tilde k}^{-1} \delta^{(3)}_{st}(\vec{\tilde k} -\vec{\tilde k}') ~.
\ee
The factor $\omega_{\tilde k}^{-1}$ is exactly the one used in \cite{vacfluctuations} (cf. Eq. (107) there). This 
is precisely the result needed to prove that if we take a DSR representing running to dimension 2 in the UV, then
its quantum vacuum fluctuations are scale-invariant. 

\section{Three point function}\label{3point}
\begin{figure}[htbp]
\begin{center}
 \includegraphics[width=0.2\textwidth]{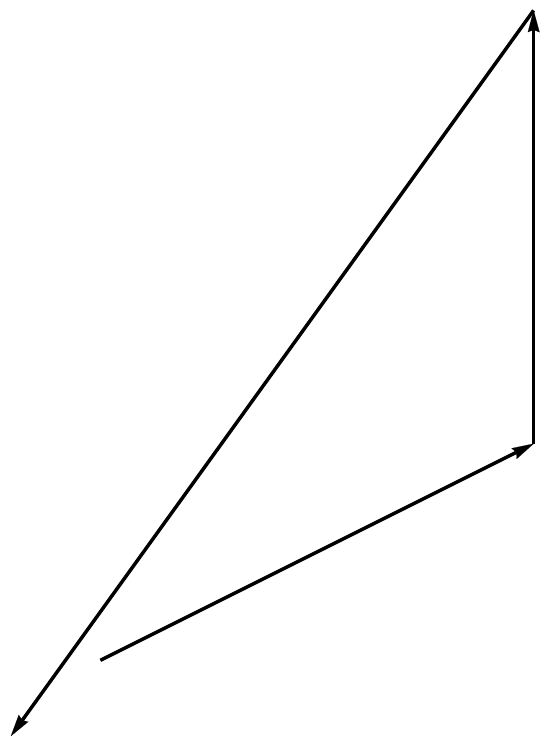}
\caption{One of the two choices of wavevectors selected by $\delta^{(3)}(\vec k_{1}\oplus \vec k_{2} \oplus \vec k_{3})$, with fixed $\vec k_{1}=(1,1/2,0)$, $\vec k_{2}=(0,1,0)$ and $\frac{1}{\kappa}=0.1$.}
\label{opentriangle}
\end{center}
\end{figure}

Novelties only arise with regards to the standard quantization procedure when we consider higher order correlators.
The three point function, for example,  has the generic form \cite{MicheleInteraction}:
\be
\langle \phi^{3}\rangle \propto \delta^{(3)}\left(\sum_{\kappa}(\vec k_{1},\vec k_{2},\vec k_{3})\right)F(\vec k_{1},\vec k_{2},\vec k_{3})
\ee
where the function $F$ depends on the exact form of the interaction. In the argument of the delta function the symbol $\sum_{\kappa}$ stands for the deformed sum of all the possible orderings of the momenta $k_{i}$ (recall that the deformed sum rule (\ref{eq:kappacomposition}) is non-abelian).
The terms appearing as arguments of the delta function can be of the four different forms:
\bea
\delta^{(3)}\left(\vec k_{1}\oplus \vec k_{2} \oplus \vec k_{3}\right)+\delta^{(3)}(\text{permutations})\label{form1}\\
\delta^{(3)}\left(\vec k_{1}\oplus (\ominus \vec k_{2}) \oplus \vec k_{3}\right)+\delta^{(3)}(\text{permutations})\label{form2}\\
\delta^{(3)}\left(\vec k_{1}\oplus (\ominus \vec k_{2}) \oplus (\ominus \vec k_{3})\right)+\delta^{(3)}(\text{permutations})\label{form3}\\
\delta^{(3)}\left(\ominus \vec k_{1}\oplus (\ominus \vec k_{2} )\oplus (\ominus \vec k_{3})\right)+\delta^{(3)}(\text{permutations}).\label{form4}
\eea
In the undeformed case (or for theories with Lorentz invariance violation) all of these options are equivalent and select wave-vectors that form a triangle (in Fourier space). The function $F$ can then make specific kinds of triangles to be dominant in the contribution to the three-point function (e.g. squeezed triangles, equilateral triangles etc.).
The situation is different for DSR. As an example let us 
consider the argument of first of the above delta functions and write its explicit form. We should take into account the fact that, as done for the two-point function, we are considering on-shell quantities, so that:
\bea
&&\vec k_{1}\oplus \vec k_{2} \oplus \vec k_{3}=\nonumber\\
&&\vec k_{1}+e^{-\omega_{k_{1}}^{\kappa}/\kappa} \vec k_{2} +e^{-(\omega_{k_{1}}^{\kappa}+\omega_{k_{2}}^{\kappa})/\kappa}  \vec k_{3}=\nonumber\\
&&\vec k_{1}+\vec k_{2}+\vec k_{3}-\lambda\left( |\vec k_{1}|\vec k_{2}+\left(|\vec k_{1}|+|\vec k_{2}|-\lambda|\vec k_{1}||\vec k_{2}|\right)\vec k_{3}\right)\nonumber\\
\eea
where in the last line we have used the on-shell condition  (\ref{eq:omega}). 
%{[\bf NB: this result is exact, NOT at leading order]}. 
We see that the Fourier modes contributing to the three-point function will no longer form a triangle. There will
 generically be a ``gap'' not allowing the triangle to close. The size of the gap depends on the momenta appearing in the delta function. In Figure \ref{opentriangle} we show one typical example of the wavevectors selected by the deformed delta function.
 
% \begin{figure}[htbp]
%\begin{center}
%\begin{subfigure}[b]{0.1\textwidth}
% \includegraphics[width=1.1\textwidth]{opentriangle1}
% \end{subfigure}
%\begin{subfigure}[b]{0.1\textwidth}
% \includegraphics[width=\textwidth]{opentriangle2}
% \end{subfigure} 
%\caption{default}
%\label{default}
%\end{center}
%\end{figure}

Similar results can be obtained using the other three kinds of delta functions listed at the beginning of this section. 
The only case in which a gap is not present is when we select (\ref{form2}) or (\ref{form3}) and set the last vector to zero. Then we do produce a collapsed triangle, i.e. a triangle in which one of the sides is a single point, since this case reduces to the two-point function studied in the previous sections:
\be
\delta(\vec k_{1}\oplus ({\ominus \vec k_{2}}) \oplus 0)=\delta(\vec k_{1}\oplus ({\ominus \vec k_{2}} ))= \delta(\vec k_{1}-\vec k_{2})e^{-\omega_{k_{1}}^{(\kappa)}/\kappa}.
\ee

The 3-point function of vacuum fluctuations in DSR therefore appears to violate translational invariance, but this is because translational invariance is in fact being enforced  in the context of a curved momentum space. This conclusion is observationally very interesting because the theory does not violate translational invariance at the level of the 2-point function, but it does appear to do so for higher order correlators only.

\section{Conclusion}
In \cite{vacfluctuations} we proposed a quantization scheme applicable to the evaluation of the power spectrum of a wide range of theories exhibiting modified dispersion relations,  with or without a preferred frame. It allowed us to {\it universally} relate dynamical dimensional reduction to 2 in the UV and the scale-invariance of vacuum quantum fluctuations. However (as was already pointed out in \cite{vacfluctuations})  the proposed quantization scheme cannot be valid for the calculation of higher order correlators in theories without a preferred frame. This is because frame-independence forces the theory to have a non-trivial multiparticle sector with non-trivial (frequently non-abelian) addition rules for energy-momentum. The power spectrum evaluation only probes the normalization of the 1-particle states; however, the higher order correlators probe the multi-particle sector, and therefore require modifications to quantization specific to frame independent theories.

In this paper we focused on the concrete example of DSR to explicitly illustrate this point. 
We showed that the rigorous quantization of DSR does reduce to the prescription proposed in \cite{vacfluctuations} for the evaluation of the 2-point function. We then presented the blueprint for evaluating the 3-point function, without specifying the concrete form of the interaction Hamiltonian.  We found the surprising result that apparent violations of translational invariance must appear in the 3-point function, which acquires the structure of delta functions on open triangles. This happens precisely because translational invariance has been enforced within the context of a curved momentum space.

The observational implications of this result are obviously interesting.  In the past there have been proposals of theories which violate isotropy~\cite{FerMag,Durrer,Tarun}, in the sense that their 2-point function has (non-vanishing) components which would otherwise be set to zero. In this paper we found that whilst this does not happen for DSR at the level of the 2-point function, it may occur when we consider the 3-point function, which fails to be proportional to a delta function on closed triangles. The structure of open triangles allowed by DSR is fully defined from the geometry of de Sitter space, rendering the theory highly predictive, albeit difficult to test for practical reasons.  Besides noise issues, the fundamental hurdle to detection would be the cosmic variance of standard theories with vanishing 3-point function except for closed triangles. Projected onto the sky, DSR would manifest itself in a three point function not proportional to a Clebsh-Gordon coefficient. The intensity of the three point function would of course depend on the strength and details of the interaction Hamiltonian of the theory. A full study of these predictions is deferred to future work. 

\section*{Acknowledgments} 
We acknowledge support from the John Templeton Foundation. The work of MA was also supported by a Marie Curie Career Integration Grant within the 7th European Community Framework Programme. JM was further funded by an STFC consolidated grant and the Leverhulme Trust.

%%%%%%%%%%%%%%%%%%%%


\begin{thebibliography}{99}

\bibitem{AmelinoCamelia:2008qg}
  G.~Amelino-Camelia,
  %``Quantum-Spacetime Phenomenology,''
  Living Rev.\ Rel.\  {\bf 16} (2013) 5
  [arXiv:0806.0339 [gr-qc]].
  %%CITATION = ARXIV:0806.0339;%%
  %174 citations counted in INSPIRE as of 26 Oct 2015
  
  \bibitem{AmelinoCamelia:2011pe}
  G.~Amelino-Camelia, L.~Freidel, J.~Kowalski-Glikman and L.~Smolin,
  %``Relative locality: A deepening of the relativity principle,''
  Gen.\ Rel.\ Grav.\  {\bf 43} (2011) 2547
   [Int.\ J.\ Mod.\ Phys.\ D {\bf 20} (2011) 2867]
  [arXiv:1106.0313 [hep-th]].
  %%CITATION = ARXIV:1106.0313;%%
  %60 citations counted in INSPIRE as of 26 Oct 2015
  
  \bibitem{AmelinoCamelia:2011bm}
  G.~Amelino-Camelia, L.~Freidel, J.~Kowalski-Glikman and L.~Smolin,
  %``The principle of relative locality,''
  Phys.\ Rev.\ D {\bf 84} (2011) 084010
  [arXiv:1101.0931 [hep-th]].
  %%CITATION = ARXIV:1101.0931;%%
  %143 citations counted in INSPIRE as of 26 Oct 2015
  
  \bibitem{Matschull:1997du}
  H.~J.~Matschull and M.~Welling,
  %``Quantum mechanics of a point particle in (2+1)-dimensional gravity,''
  Class.\ Quant.\ Grav.\  {\bf 15} (1998) 2981
  [gr-qc/9708054].
  %%CITATION = GR-QC/9708054;%%
  %96 citations counted in INSPIRE as of 26 Oct 2015
  
  \bibitem{Freidel:2005me}
  L.~Freidel and E.~R.~Livine,
  %``Effective 3-D quantum gravity and non-commutative quantum field theory,''
  Phys.\ Rev.\ Lett.\  {\bf 96} (2006) 221301
  [hep-th/0512113].
  %%CITATION = HEP-TH/0512113;%%
  %174 citations counted in INSPIRE as of 26 Oct 2015

\bibitem{Arzano:2014ppa}
  M.~Arzano, D.~Latini and M.~Lotito,
  %``Group Momentum Space and Hopf Algebra Symmetries of Point Particles Coupled to 2+1 Gravity,''
  SIGMA {\bf 10} (2014) 079
  [arXiv:1403.3038 [gr-qc]].
  %%CITATION = ARXIV:1403.3038;%%
  %5 citations counted in INSPIRE as of 26 Oct 2015
  
\bibitem{Gubitosi:2013rna}
  G.~Gubitosi and F.~Mercati,
  %``Relative Locality in $\kappa$-Poincar\'e,''
  Class.\ Quant.\ Grav.\  {\bf 30} (2013) 145002
  [arXiv:1106.5710 [gr-qc]].
  %%CITATION = ARXIV:1106.5710;%%
  %29 citations counted in INSPIRE as of 26 Oct 2015
  
\bibitem{vacfluctuations}
  M.~Arzano, G.~Gubitosi, J.~Magueijo and G.~Amelino-Camelia,
  %``Vacuum fluctuations in theories with deformed dispersion relations,''
  Phys.\ Rev.\ D {\bf 91} (2015) 12,  125031
  [arXiv:1505.05021 [gr-qc]].
  %%CITATION = ARXIV:1505.05021;%%
  %1 citations counted in INSPIRE as of 15 Oct 2015
  
  
    \bibitem{measure}
  G.~Amelino-Camelia, M.~Arzano, G.~Gubitosi and J.~Magueijo,
  %``Dimensional reduction in momentum space and scale-invariant cosmological fluctuations,''
  Phys.\ Rev.\ D {\bf 88} (2013) 10,  103524
  [arXiv:1309.3999 [gr-qc]].
  %%CITATION = ARXIV:1309.3999;%%
  %10 citations counted in INSPIRE as of 15 Oct 2015
  
  \bibitem{essay}
  G.~Amelino-Camelia, M.~Arzano, G.~Gubitosi and J.~Magueijo,
  %``Gravity as the breakdown of conformal invariance,''
  arXiv:1505.04649 [gr-qc].
  %%CITATION = ARXIV:1505.04649;%%
  %3 citations counted in INSPIRE as of 15 Oct 2015
  
  

  
 \bibitem{arzano-marciano}
  M.~Arzano and A.~Marciano,
  %``Fock space, quantum fields and kappa-Poincare symmetries,''
  Phys.\ Rev.\ D {\bf 76}, 125005 (2007)
  [arXiv:0707.1329 [hep-th]].

\bibitem{Arzano:2013sta} 
  M.~Arzano, J.~Kowalski-Glikman and T.~Trzesniewski,
  %``Beyond Fock space in three dimensional semiclassical gravity,''
  Class.\ Quant.\ Grav.\  {\bf 31}, no. 3, 035013 (2014)
  [arXiv:1305.6220 [hep-th]].
  
  %\cite{Majid:1994cy}
\bibitem{Majid:1994cy}
  S.~Majid and H.~Ruegg,
  %``Bicrossproduct structure of kappa Poincare group and noncommutative geometry,''
  Phys.\ Lett.\ B {\bf 334} (1994) 348
  [hep-th/9405107].
  %%CITATION = HEP-TH/9405107;%%
  %466 citations counted in INSPIRE as of 15 Oct 2015
  
\bibitem{MicheleQuantization}
  M.~Arzano,
  %``Anatomy of a deformed symmetry: Field quantization on curved momentum space,''
  Phys.\ Rev.\ D {\bf 83} (2011) 025025
  [arXiv:1009.1097 [hep-th]].
  %%CITATION = ARXIV:1009.1097;%%
  %14 citations counted in INSPIRE as of 06 mar 2015


\bibitem{MicheleInteraction}
  G.~Amelino-Camelia and M.~Arzano,
  %``Coproduct and star product in field theories on Lie algebra noncommutative space-times,''
  Phys.\ Rev.\ D {\bf 65}, 084044 (2002)
  [hep-th/0105120].


\bibitem{FerMag} 
  P.~G.~Ferreira and J.~Magueijo,
  %``The Closet nonGaussianity of anisotropic Gaussian fluctuations,''
  Phys.\ Rev.\ D {\bf 56}, 4578 (1997)
  [astro-ph/9704052].

\bibitem{Durrer} 
  R.~Durrer, T.~Kahniashvili and A.~Yates,
  %``Microwave background anisotropies from Alfven waves,''
  Phys.\ Rev.\ D {\bf 58}, 123004 (1998)
  [astro-ph/9807089].

\bibitem{Tarun} 
  A.~Hajian and T.~Souradeep,
  %``Measuring statistical isotropy of the CMB anisotropy,''
  Astrophys.\ J.\  {\bf 597}, L5 (2003)
  [astro-ph/0308001].


\end{thebibliography}
\end{document}